\documentstyle[twoside,epsf]{article}

\catcode`\@=11 \long\def\@makefntext#1{ \protect\noindent \hbox to
3.2pt {\hskip-.9pt
$^{{\eightrm\@thefnmark}}$\hfil}#1\hfill}       

\def\@makefnmark{\hbox to 0pt{$^{\@thefnmark}$\hss}}    

\def\ps@myheadings{\let\@mkboth\@gobbletwo
\def\@oddhead{\hbox{}
\rightmark\hfil\eightrm\thepage}
\def\@oddfoot{}\def\@evenhead{\eightrm\thepage\hfil
\leftmark\hbox{}}\def\@evenfoot{}
\def\sectionmark##1{}\def\subsectionmark##1{}}

\newcommand{\be}{\begin{equation}}
\newcommand{\ee}{\end{equation}}
\newcommand{\ba}{\begin{eqnarray}}
\newcommand{\ea}{\end{eqnarray}}
\newcommand{\ban}{\begin{eqnarray*}}
\newcommand{\ean}{\end{eqnarray*}}

\newcommand{\braket}[2]{\mbox{$ \langle #1 | #2 \rangle $}}
\newcommand{\sandwich}[3]{\mbox{$ \langle #1 | #2 | #3 \rangle $}}
\newcommand{\ket}[1]{\mbox{$ | #1 \rangle $}}
\newcommand{\bra}[1]{\mbox{$ \langle #1 | $}}

\newcommand{\si}{\sigma}
\newcommand{\demi}{\frac{1}{2}}
\newcommand{\compl}{\begin{picture}(8,8)\put(0,0){C}\put(3,0.3){\line(0,1){7}}\end{picture}}

\newcommand{\one}{\leavevmode\hbox{\small1\normalsize\kern-.33em1}}

\oddsidemargin=\evensidemargin 
\newcounter{sectionc}\newcounter{subsectionc}\newcounter{subsubsectionc}
\renewcommand{\section}[1] {\vspace{12pt}\addtocounter{sectionc}{1}
\setcounter{subsectionc}{0}\setcounter{subsubsectionc}{0}\noindent
    {\tenbf\thesectionc. #1}\par\vspace{5pt}}
\renewcommand{\subsection}[1] {\vspace{12pt}\addtocounter{subsectionc}{1}
\setcounter{subsubsectionc}{0}\noindent
{\bf\thesectionc.\thesubsectionc. {\kern1pt \bfit
#1}}\par\vspace{5pt}}
\renewcommand{\subsubsection}[1] {\vspace{12pt}\addtocounter{subsubsectionc}{1}
    \noindent{\tenrm\thesectionc.\thesubsectionc.\thesubsubsectionc.
    {\kern1pt \tenit #1}}\par\vspace{5pt}}

\newcounter{appendixc}
\newcounter{subappendixc}[appendixc]
\newcounter{subsubappendixc}[subappendixc]
\renewcommand{\thesubappendixc}{\Alph{appendixc}.\arabic{subappendixc}}
\renewcommand{\thesubsubappendixc}
    {\Alph{appendixc}.\arabic{subappendixc}.\arabic{subsubappendixc}}

\renewcommand{\appendix}[1] {\vspace{12pt}
        \refstepcounter{appendixc}
        \setcounter{figure}{0}
        \setcounter{table}{0}
        \setcounter{lemma}{0}
        \setcounter{theorem}{0}
        \setcounter{corollary}{0}
        \setcounter{definition}{0}
        \setcounter{equation}{0}
        \renewcommand{\thefigure}{\Alph{appendixc}.\arabic{figure}}
        \renewcommand{\thetable}{\Alph{appendixc}.\arabic{table}}
        \renewcommand{\theappendixc}{\Alph{appendixc}}
        \renewcommand{\thelemma}{\Alph{appendixc}.\arabic{lemma}}
        \renewcommand{\thetheorem}{\Alph{appendixc}.\arabic{theorem}}
        \renewcommand{\thedefinition}{\Alph{appendixc}.\arabic{definition}}
        \renewcommand{\thecorollary}{\Alph{appendixc}.\arabic{corollary}}
        \renewcommand{\theequation}{\Alph{appendixc}.\arabic{equation}}
        \noindent{\tenbf Appendix \theappendixc #1}\par\vspace{5pt}}
\newcommand{\subappendix}[1] {\vspace{12pt}
        \refstepcounter{subappendixc}
        \noindent{\bf Appendix \thesubappendixc. {\kern1pt \bfit #1}}
    \par\vspace{5pt}}
\newcommand{\subsubappendix}[1] {\vspace{12pt}
        \refstepcounter{subsubappendixc}
        \noindent{\rm Appendix \thesubsubappendixc. {\kern1pt \tenit #1}}
    \par\vspace{5pt}}

\newcommand{\textlineskip}{\baselineskip=13pt}
\newcommand{\smalllineskip}{\baselineskip=10pt}

\newcommand{\copyrightheading}[1]
    {\vspace*{-2.5cm}\smalllineskip{\flushleft
    {\footnotesize Quantum Information and Computation, Vol.~1, No.~0 (2001) 000--000 #1}\\
    {\footnotesize \copyright\kern2pt Rinton Press}\\
     }}

\newcommand{\publisher}[2]{{\begin{center}\footnotesize\smalllineskip
    Received #1\\
    Revised #2
    \end{center}
    }}

\def\abstracts#1#2#3{{
    \centering{\begin{minipage}{4.5in}\footnotesize\baselineskip=10pt
    \parindent=0pt #1\par
    \parindent=15pt #2\par
    \parindent=15pt #3
    \end{minipage}}\par}}

\def\keywords#1{{
    \centering{\begin{minipage}{4.5in}\footnotesize\baselineskip=10pt
    {\footnotesize\it Keywords}\/: #1
     \end{minipage}}\par}}
\def\communicate#1{{
    \centering{\begin{minipage}{4.5in}\footnotesize\baselineskip=10pt
    {\footnotesize\it Communicated by}\/: #1
     \end{minipage}}\par}}

\renewenvironment{thebibliography}[1]
        {\frenchspacing
     \ninerm\baselineskip=11pt
         \begin{list}{\arabic{enumi}.}
        {\usecounter{enumi}\setlength{\parsep}{0pt}
     \setlength{\leftmargin 12.7pt}{\rightmargin 0pt}
         \setlength{\itemsep}{0pt} \settowidth
    {\labelwidth}{#1.}\sloppy}}{\end{list}}

\newcounter{itemlistc}
\newcounter{romanlistc}
\newcounter{alphlistc}
\newcounter{arabiclistc}

\newcommand{\fcaption}[1]{
        \refstepcounter{figure}
        \setbox\@tempboxa = \hbox{\footnotesize Fig.~\thefigure. #1}
        \ifdim \wd\@tempboxa > 5in
           {\begin{center}
        \parbox{5in}{\footnotesize\smalllineskip Fig.~\thefigure. #1}
            \end{center}}
        \else
             {\begin{center}
             {\footnotesize Fig.~\thefigure. #1}
              \end{center}}
        \fi}

\newcommand{\tcaption}[1]{
        \refstepcounter{table}
        \setbox\@tempboxa = \hbox{\footnotesize Table~\thetable. #1}
        \ifdim \wd\@tempboxa > 5in
           {\begin{center}
        \parbox{5in}{\footnotesize\smalllineskip Table~\thetable. #1}
            \end{center}}
        \else
             {\begin{center}
             {\footnotesize Table~\thetable. #1}
              \end{center}}
        \fi}

\def\pmb#1{\setbox0=\hbox{#1}
    \kern-.025em\copy0\kern-\wd0
    \kern.05em\copy0\kern-\wd0
    \kern-.025em\raise.0433em\box0}

\def\fnt#1#2{\footnotetext{\kern-.3em
    {$^{\mbox{\scriptsize #1}}$}{#2}}}

\def\fpage#1{\begingroup
\voffset=.3in
\thispagestyle{empty}\begin{table}[b]\centerline{\footnotesize #1}
    \end{table}\endgroup}

\def\runninghead#1#2{\pagestyle{myheadings}
\markboth{{\protect\footnotesize\it{\quad #1}}\hfill}
{\hfill{\protect\footnotesize\it{#2\quad}}}} \headsep=15pt

\font\tenrm=cmr10 \font\tenit=cmti10 \font\tenbf=cmbx10
\font\bfit=cmbxti10 at 10pt \font\ninerm=cmr9 
 \font\eightrm=cmr8

\def\FigName{figure}%
\newbox\captionbox
\long\def\@makecaption#1#2{%
  \ifx\FigName\@captype
    \vskip\abovecaptionskip
    \setbox\tempbox\hbox{{\figurecaptionfont #1\hskip1em #2}}
    \ifdim\wd\tempbox< 28pc
    \centerline{\box\tempbox}
    \else
    {\figurecaptionfont #1\hskip1em #2\par}
\fi\else
    \setbox\tempbox\hbox{{\tablecaptionfont #1\hskip1em #2}}
    \ifdim\wd\tempbox< 28pc
    \centerline{\box\tempbox}
    \else
    {\tablecaptionfont #1\hskip1em #2\par}%
    \fi
 \vskip\belowcaptionskip
 \fi}
\InputIfFileExists{psfig.sty}
{\typeout{^^Jpsfig.sty inputed...ok}}{\typeout{^^JWarning:
psfig.sty could be be found.^^J}}
\InputIfFileExists{epsfsafe.tex}
{\typeout{^^Jepsfsafe.tex inputed...ok}}
            {\typeout{^^JWarning: epsfsafe.tex could not be found.^^J}}
\InputIfFileExists{epsfig.sty}
{\typeout{^^Jepsfig.sty inputed...ok}}{\typeout{^^JWarning:
epsfig.sty could not be found.^^J}}
\InputIfFileExists{epsf.sty}
{\typeout{^^Jepsf.sty inputed...ok}}{\typeout{^^JWarning: epsf.sty could not be found.^^J}}%
\def\fps@figure{tbp}
\def\ftype@figure{1}
\def\ext@figure{lof}
\def\fnum@figure{Fig.\ \thefigure}
\def\qed{\hbox{${\vcenter{\vbox{              
   \hrule height 0.4pt\hbox{\vrule width 0.4pt height 6pt
   \kern5pt\vrule width 0.4pt}\hrule height 0.4pt}}}$}}

\begin{document}
\setlength{\textheight}{8.0truein}    

\runninghead{Title   $\ldots$}
            {Author(s) $\ldots$}

\normalsize\textlineskip \thispagestyle{empty}
\setcounter{page}{1}

\copyrightheading{} 

\vspace*{0.88truein}

\fpage{1} \centerline{\bf
SECURITY BOUNDS IN QUANTUM CRYPTOGRAPHY } \vspace*{0.035truein}
\centerline{\bf USING $d$-LEVEL SYSTEMS} \vspace*{0.37truein}
\centerline{\footnotesize
ANTONIO ACIN\footnote{Presently at the Institute of Photonic
Sciences, 29 Jordi Girona, 08034 Barcelona, Spain.
}}\vspace*{0.015truein} \centerline{\footnotesize\it GAP-Optique,
University of Geneva, 20, Rue de l'\'Ecole de M\'edecine}
\baselineskip=10pt \centerline{\footnotesize\it CH-1211 Geneva,
Switzerland} \vspace*{10pt} \centerline{\footnotesize NICOLAS
GISIN}\vspace*{0.015truein} \centerline{\footnotesize\it
GAP-Optique, University of Geneva, 20, Rue de l'\'Ecole de
M\'edecine} \baselineskip=10pt \centerline{\footnotesize\it
CH-1211 Geneva, Switzerland} \vspace*{10pt}
\centerline{\footnotesize VALERIO SCARANI} \vspace*{0.015truein}
\centerline{\footnotesize\it GAP-Optique, University of Geneva,
20, Rue de l'\'Ecole de M\'edecine} \baselineskip=10pt
\centerline{\footnotesize\it CH-1211 Geneva, Switzerland}
\vspace*{0.225truein} \publisher{(received date)}{(revised date)}

\vspace*{0.21truein} \abstracts{
We analyze the security of quantum cryptography schemes for
$d$-level systems using 2 or $d+1$ maximally conjugated bases,
under individual eavesdropping attacks based on cloning machines
and measurement after the basis reconciliation. We consider
classical advantage distillation protocols, that allow to extract
a key even in situations where the mutual information between the
honest parties is smaller than the eavesdropper's information. In
this scenario, advantage distillation protocols are shown to be as
powerful as quantum distillation: key distillation is possible
using classical techniques if and only if the corresponding state
in the entanglement based protocol is distillable.}{}{}

\vspace*{10pt} \keywords{Quantum Cryptography, Key Distillation,
Quantum Distillation} \vspace*{3pt} \communicate{to be filled by
the Editorial}

\vspace*{1pt}\textlineskip  
\section{Introduction}           
\vspace*{-0.5pt} \noindent
Quantum Cryptography (QC) is a physically secure protocol to
distribute a secret key between two authorized partners, Alice and
Bob, at distant locations \cite{review}. Its security is based on
the no-cloning theorem: if Alice encodes the correlation in the
state of a $d$-dimensional quantum system ({\em qudit}) that she
sends to Bob, an eavesdropper Eve cannot extract any information
without introducing errors. By estimating {\em a posteriori} the
errors in their correlations, Alice and Bob can detect the
presence of the spy on the line. Of course, zero error can never
be achieved in practice, even in the absence of Eve. By
continuity, if the error is ``small" one expects that it will
still be possible to extract a secret key from the noisy data
\cite{uncond}. At the other extreme, if the error is large, then
Eve could have obtained ``too much" information, so the only way
for Alice and Bob to guarantee security is to stop the protocol
and wait for better times. It becomes then important to quantify
the amount of error that can be tolerated on the Alice-Bob
channel: this value measures the {\em robustness} of a QC
protocol.

The problem of the extraction of a secret key from noisy data is
of course not specific of quantum key distribution (QKD). In a
typical cryptography scenario, Alice, Bob and Eve share $N$
independent realizations of a triple $(a,b,e)$ of {\em classical}
random variables, distributed according to some probability law,
$P(A,B,E)$. The variables $a$ and $b$ are both $d$-valued, we say
that Alice and Bob encode their information in {\em dits}. Eve can
always process her data to obtain the optimal guesses for the
values of $a$ and $b$, $e_a,e_b$, with $e_x$ the $d$-valued guess
for $x$. From $P$, one can in particular calculate the mutual
information: \ba
I(A:B)&=&H(A)+H(B)-H(AB)\,,\\
I(A:E)&=&H(A)+H(E_A)-H(AE_A)\,,\\
I(B:E)&=&H(B)+H(E_B)-H(BE_B)\,, \ea where $H$ is the Shannon
entropy, measured in dits, e.g. $H(A)=-\sum_{k=0}^{d-1}P(a=k)
\log_d P(a=k)$.

To {\em extract a secret key} from the raw data means that Alice
and Bob are able to process their data and communicate classically
in order to end with $n<N$ realizations of new variables
$(a',b',e')$ such that asymptotically $I(A':B')=1$, and
$I(A':E')=I(B':E')=0$. In other words, the processed variables
must be distributed according to a probability law $P'$ of the
form $P'(A',B')P'(E)$, with $P'(a'=b')=1$. To date, no necessary
and sufficient criterion is known to decide whether a secret key
can be extracted from a given classical distribution $P(A,B,E)$.
Basically two results are known:

{\em CK criterion.} If $I(A:B)>I_E=\min[I(A:E),I(B:E)]$, then a
secret key of length $n = [I(A:B)-I_E]\,N$ can be extracted using
one-way classical data processing. This theorem, given by
Csisz\'ar and K\"{o}rner in 1978 \cite{csi}, formalizes the
intuitive idea that if Eve has less information than Bob on
Alice's string (or, than Alice on Bob's string), the extraction of
a secret key is possible. It consists of the following two steps:
error correction followed by privacy amplification \cite{BBCM}.
The whole process is done using unidirectional communication.

{\em AD criterion.} Even if $I(A:B)\leq I_E$ however, in some
cases a secret key between Alice and Bob can be extracted. This is
because (i) Eve has made some errors, her information is bounded,
and (ii) Alice and Bob share a classical authenticated and
error-free channel: in other words, Eve can listen to the
classical communication but can neither modify nor even disturb
it. These protocols were introduced in 1993 by Maurer
\cite{maurer}, who called them {\em advantage distillation
protocols}. They require two-way communication between Alice and
Bob and are rather inefficient. Very little is known about the
conditions (for instance, in terms of Eve's error probability or
information) such that a key can be distilled using these
protocols.

Most of the works of QC define robustness by using CK. AD
protocols in QC were considered a few years ago by Gisin and Wolf
\cite{wolf}, who studied the case of qubit encoding ($d=2$). In
this paper, we analyze QC protocols with $d$-level quantum states
or qudits \cite{cerf} under individual attacks based on cloning
machines. In Section 2, we describe our scenario: the protocols
and the individual attacks considered. We also present the
entanglement based version of all these protocols. Indeed,
although entanglement is in principle not required for a secure
key distribution, it is known that any QKD protocol can be easily
translated into an analogous entanglement based protocol. In
Section 3, we generalize the result of Gisin and Wolf to the case
of qudits: we show that, under our assumptions, classical
advantage distillation works for $d$-level protocols if and only
if the quantum state shared by Alice and Bob before the
measurement in the corresponding entanglement based protocol is
entangled and distillable. In Section 4, we discuss the link
between the CK criterion and the violation of Bell's inequalities,
noticed for qubits in Refs \cite{bell1,fuchs}. Section 5 is a
conclusion, in which we review some interesting open questions.

\section{QC with qudits}
\noindent
\subsection{The protocol}
\noindent A general scheme for QC with qudits, generalizing BB84
protocol for qubits \cite{BB84}, has been presented by Cerf et al.
\cite{cerf}. Central to this development is the notion of mutually
unbiased bases: two bases $B_1=\big\{\ket{k}\big\}$ and
$B_2=\big\{\ket{\bar{l}}\big\}$ are called unbiased (or maximally
conjugated) if $|\braket{k}{\bar{l}}|^2=\frac{1}{d}$ for all
vectors in each basis. For qudits, one can find at most $d+1$
maximally conjugated bases \cite{wootters}. Once a computational
basis $B_1=\big\{\ket{0},\ket{1},...,\ket{d-1}\big\}$ is
arbitrarily chosen, one can always construct at least one unbiased
basis, the so-called {\em Fourier-dual basis} \ba
\ket{\bar{l}}&=&\frac{1}{\sqrt{d}}\,\sum_{k=0}^{d-1}e^{2\pi i
kl/d}\ket{k}\,. \ea

Let ${\cal{B}}=\big\{B_1,...,B_n\big\}$, with $2\leq n\leq d+1$, a
set of $n$ mutually unbiased bases, where $B_1$ is chosen as the
computational basis. Alice prepares at random one state belonging
to one of these bases and sends it to Bob. Bob receives the qudit,
and measures it in one of the bases of the set ${\cal{B}}$. Then,
(i) if Alice and Bob use the same basis, their results are
perfectly correlated; (ii) if they use different bases, their
results are totally uncorrelated. Later, they reveal publicly the
basis that they used: they keep the items where they used the same
basis and discard the others. So, after this sifting procedure,
Alice and Bob are left with a fraction $\frac{1}{n}$ of the raw
list. In the absence of any disturbance, and in particular in the
absence of Eve, these dits are perfectly correlated.

It is straightforward to construct the corresponding entanglement
based protocol \cite{BBM,noteekert}. Alice prepares a maximally
entangled state \ba \ket{\Phi}&=&\frac{1}{\sqrt{d}}\,
\sum_{k=0}^{d-1}\ket{k}_A\ket{k}_B\,, \ea keeps one qudit and
sends the other to Bob. The maximally entangled state is maximally
correlated in all the bases, since for all unitary operations
$U\in SU(d)$, \ba (U\otimes U^*)\ket{\Phi}&=&\ket{\Phi}\,. \ea
After the state distribution, Alice and Bob measure at random in
one of the bases of ${\cal B}$ (more precisely Bob's set of bases
is ${\cal B}^*$). They announce the measurement bases. Only those
symbols where they chose the same basis are accepted, giving a
list of perfectly correlated dits. Note that Alice's measurement
outcome is completely equivalent to the previous state
preparation.

For the rest of the article, and for consistency in the
presentation, we will mainly concentrate on entanglement based
protocols. But it has to be stressed that some of the ideas are
especially meaningful for protocols without entanglement. For
instance, whenever we speak about classical key distillation
protocols, we also refer to protocols without entanglement.


\subsection{Generalities about Eve's attacks}
\noindent Now we must study Eve's attacks on the qudits travelling
to Bob. To find the most general eavesdropping attack for a QC
protocol is a very hard problem. In this article we restrict our
considerations to {\em individual attacks}: first, Eve lets the
incoming qudit interact in a suitable way with some auxiliary
quantum system she has prepared in a reference state $\ket{R}$.
Then she lets the qudit go to Bob and stores her system. When
Alice reveals the bases, Eve performs the measurement that allows
her to gain some information about the qudit. Note that: (i) no
coherent attacks will be considered, (ii) Eve is supposed to
measure her system after the basis reconciliation and (iii) the
individual attack does not change from symbol to symbol
\cite{BBCM}.

Thus, after Eve's intervention, the total quantum state reads \ba
\ket{\Psi}_{ABE}&=& \big(\one_A\otimes U_{BE}\big)\,
\ket{\Phi}_{AB}\otimes\ket{R}_E\,. \ea
Since Eve does not modify the local
density matrix $\rho_A=\frac{1}{d}\one$ of Alice, we have $H(A)=1$ . 
We also focus on attacks such that Eve introduces the same amount
of error in all bases: $P(a\neq b|B_i)\equiv D$ for all
$i=1,...,j$. Indeed, it was proven in \cite{CG} that, given an
asymmetric eavesdropping strategy, one can always design a
symmetric attack as powerful as it.
The mutual information Alice-Bob is thus simply \ba I(A:B)&=& 1-
H(\{D,1-D\})\,.\ea To go further, one must find Eve's {\em optimal
individual attack}. Since Eve can gain more information by
introducing larger errors, it is natural to optimize Eve's attack
conditioned to a fixed amount of error $D$ in the correlations
Alice-Bob. This implies that, after optimization, $P(A,B,E)$ is
ultimately only a function of $D$, and the condition for Alice and
Bob to extract a secret key will be of the form $D<\bar{D}$, for a
bound $\bar{D}$ to be calculated. 
If Alice and Bob find $D\geq\bar{D}$, they simply stop the
protocol. Therefore, the value of $\bar{D}$ does not quantify the
security, but the {\em robustness} of the protocol. If $\bar D$
turns out to be very small, the QKD protocol is not practical.
According to whether we use the CK or the AD criterion to quantify
the robustness, we shall find two different {\em robustness
bounds}, $D^{CK}$ and $D^{AD}$, with of course $D^{CK}\leq
D^{AD}$.

The question is: which quantity should the individual attack
``optimize"? It is commonly accepted that we must {\em maximize
the mutual information} Alice-Eve $I(A:E)$ and/or Bob-Eve $I(B:E)$
--- it will turn out that the optimal incoherent eavesdropping
yields $I(A:E)=I(B:E)$. We follow this definition, although, as
one of the conclusions of this work, it will be stressed that
different optimizations are worth exploring. Even if now, with all
our assumptions, the problem of finding Eve's attack is formulated
in a more precise way, the optimal attack is still not easy to
find. We analyze the individual attacks based on cloning machines
given in Ref. \cite{cerf}. These individual attacks are proven to
be optimal for $d=2$, with two \cite{fuchs} and three bases
\cite{sixstate}, and $d=3$ and four bases \cite{bruss}. For larger
$d$, they are optimal under the assumption that Eve's best
strategy consists of using one of the cloning machines described
in \cite{clon}; this assumption seems plausible but has not been
proven. The next subsection describes these attacks.

\subsection{Cloning machine eavesdropping}
\label{optatt}  \noindent Following Cerf et al. \cite{cerf}, we
consider only 2-bases protocols, choosing the two basis as
Fourier-dual of one another, and $(d+1)$-bases protocols
\cite{exist}. These are the natural generalizations, respectively,
of the BB84 \cite{BB84} and of the six-state \cite{sixstate}
protocols for two qubits.

The evolution induced by Eve's action is built using the {\em
cloning machines} introduced in Ref. \cite{clon}. The reference
state for Eve is the maximally entangled state of two qudits,
$\ket{R}\equiv\ket{\Phi}$. The initial state
$\ket{\Phi}_{AB}\ket{\Phi}_{E_1E_2}$ is sent onto \ba
\ket{\Psi}_{ABE_1E_2}& = & \sum_{m,n=0}^{d-1}a_{m,n} \,
U_{m,n}^{(B)}\ket{\Phi}_{AB}\, U_{m,-n}^{(E_2)}\ket{\Phi}_{E_1E_2}
\label{clone} \ea where $U_{m,n}$ is the unitary operation that
acts on the computational basis as \ba U_{m,n}\ket{k}&=& e^{2\pi i
kn/d}\ket{(k+m)\,\mbox{mod}\,d}\,. \label{umn}\ea In other words,
$U_{mn}$ introduces a phase shift measured by $n$ and an index
shift measured by $m$. $U_{m,n}^{(B)}$ and $U_{m,-n}^{(E_2)}$
indicate that these transformations apply to Bob's and Eve's
second system. The coefficients $a_{m,n}$ are determined by
imposing the requirements discussed above (same amount of errors
for all bases), and then optimizing Eve's information for a given
error $D$. The detailed calculation of these coefficients can be
found in \cite{cerf}. Writing $F=1-D$, the fidelity of the
cryptography protocol, one finds for the 2-bases protocol, \ba
\begin{array}{lll} a_{0,0}&=\,F\,;\\a_{m,0}=a_{0,n}&\equiv\, x=
\sqrt{\frac{F(1-F)}{d-1}}&\mbox{ for }m,n\neq 0\,;\\a_{m,n}&\equiv
y =\frac{1-F}{d-1}& \mbox{ for }m,n\neq 0\,.\end{array}
\label{as2}\ea For the $(d+1)$-bases protocol, one finds \ba
\begin{array}{lll} a_{0,0}&\equiv v
=\sqrt{\frac{(d+1)F-1}{d}}\,;\\a_{m,n}&\equiv z =
\sqrt{\frac{1-F}{d(d-1)}}&\mbox{ for $m\neq 0$ or $n\neq 0$.}
 \end{array}\label{asd1}\ea
Note that the states $\ket{B_{m,n}}=[\one\otimes
U_{m,n}]\ket{\Phi}$ are mutually orthogonal --- in fact, they form
a basis of maximally entangled states of two qudits. In particular
then \ba \rho_{AB}(F)&=& \sum_{m,n=0}^{d-1}|a_{m,n}(F)|^2 \,
\ket{B_{m,n}}\bra{B_{m,n}}\,. \label{rhoab}\ea

The transformation defined by (\ref{clone}) can be seen as a
cloning machine, where Bob's state is the state to be copied, the
first qudit of Eve, $E_1$, is Eve's clone, and her second qudit
$E_2$ is the ancilla. After this interaction Eve waits for the
basis reconciliation. Once the used basis has been announced, Eve
can gain partial information about Alice's and Bob's symbols by
measuring her two qudits. We will consider the measurements
discussed in Ref. \cite{cerf} for both 2-bases and $(d+1)$-bases
protocols that maximize Eve's information. These measurements also
minimize Eve's error probability and are an example of the
so-called square-root measurements \cite{chefles}. It turns out
that (i) the measurement on $E_1$ gives the estimate $e_a$ for
Alice's dit; (ii) the measurement on $E_2$ gives deterministically
the value of the error introduced on Bob's side, $\chi=b-a$. Since
Eve deterministically knows the difference between Alice's and
Bob's symbols, she has $I_{AE}=I_{BE}$.

We have presently collected all the tools we need to study the
robustness bounds $D^{AD}$ (Section 3) and $D^{CK}$ (Section 4) on
QC protocols with entangled qudits.

\section{Advantage distillation and distillation of entanglement}
\noindent In this Section, we prove
the following

Theorem: {\em Let $D^{AD}$ and $D^{ED}$ denote the two bounds: (i)
a secret key can be extracted by advantage distillation for
$D<D^{AD}$, and (ii) $\rho_{AB}(F)$ is distillable for
$D=1-F<D^{ED}$. Then, for any $d$, and for both the 2-bases and
the the $(d+1)$-bases protocols,} \ba
D^{AD}&=&D^{ED}\,.\label{thm1}\ea In words: advantage distillation
protocols can be used to extract a secret key if and only if the
state $\rho_{AB}$ (\ref{rhoab}), obtained after the cloning based
attack, is entangled and distillable.

Actually, we have rigorous proofs for the $d+1$-bases protocols
for all dimension and for the 2-bases protocols up to $d=15$. For
two bases and $d>15$ the validity of the theorem is conjectured.

\begin{figure} [htbp]
\vspace*{13pt}
\centerline{\psfig{file=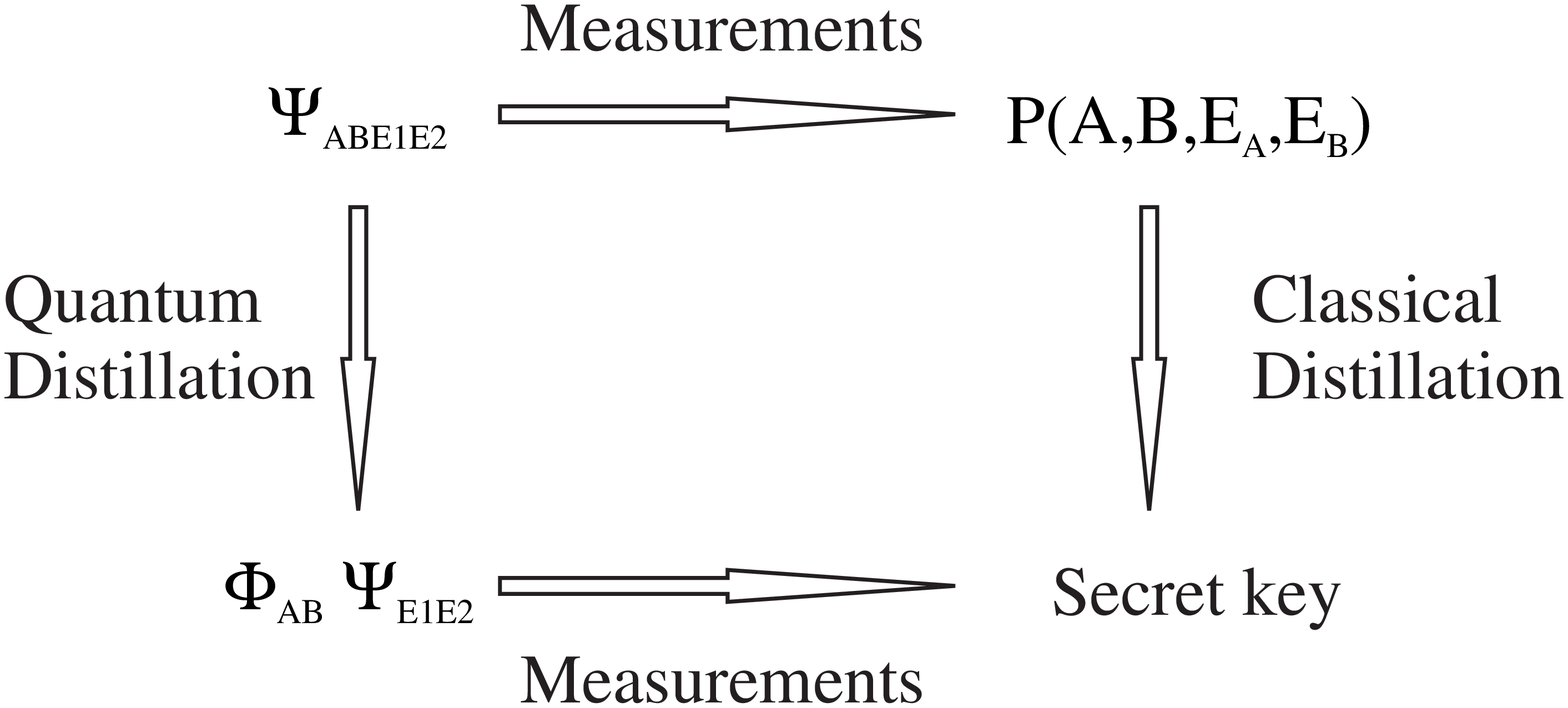, width=8.2cm}} 
\vspace*{13pt} \fcaption{\label{figcomm} Diagram illustrating the
meaning of (\ref{thm1}): the two protocols ``measure the state,
then apply advantage distillation" and ``distill the entanglement,
then measure the state" work up to the same amount of error in the
correlations Alice-Bob.}
\end{figure}

The meaning of this result is schematized in Fig. \ref{figcomm}.
We start with a quantum state $\ket{\Psi}_{ABE}$, and want to end
up with a probability distribution $P(A,B)P(E)$ with $P(a=b)=1$.
In the Introduction, we considered the following protocol: (i) the
state is measured, giving $P(A,B,E)$ \cite{notepr}; (ii) Alice and
Bob process their classical data, using AD, to factor Eve out. Let
us again emphasize here that no entanglement is actually required
for distributing the probabilities $P(A,B,E)$. But one can as well
consider {\em quantum privacy amplification}: (i') Alice and Bob
distill a maximally entangled state $\ket{\Phi}$, and since pure
state entanglement is ``monogamous" Eve is certainly factored out;
(ii') They make the measurements on $\ket{\Phi}$, and obtain the
secret key. Our Theorem thus means that these two protocols work
up to exactly the same amount of error $\bar{D}$. In other words,
as far as robustness is concerned, there seem to be no need for
entanglement distillation in QC, one can as well process the
classical information.



The proof of the Theorem is given in two steps:

{\em Step 1} (subsection 3.1): we calculate $D^{ED}$ at which
$\rho_{AB}$ ceases to be distillable. We also prove --- for all
the $(d+1)$-bases protocols, and numerically for the 2-bases
protocol up to $d=15$ --- that $\rho_{AB}$ becomes separable at
that point, that is, for no value of $D$ the state $\rho_{AB}$ is
bound entangled.

{\em Step 2} (subsection 3.2): we construct an advantage
distillation protocol that works for all $D< D^{ED}$, so that
$D^{AD}\geq D^{ED}$.

These two steps conclude the proof of (\ref{thm1}), taking into
account the following result \cite{crypto2000}: If
$\ket{\Psi}_{ABE}$ is such that $\rho_{AB}$ is separable, then,
whatever Alice and Bob do, there exists a measurement of Eve such
that the {\em intrinsic information} Alice-Bob for the derived
probability distribution $P(A,B,E)$ \ba \label{intrinf}
I(A:B\downarrow E)&=&\inf_{E\rightarrow \bar{E}}I(A:B|\bar{E}) \ea
goes to zero. In fact, the vanishing of the intrinsic information
implies that no secret key can be extracted \cite{crypto2000}.
Since for $D=D^{AD}$ the quantum state shared by Alice and Bob is
separable, Eve can simply apply this measurement preventing Alice
and Bob to establish a key.

One may wonder whether, at this critical point, the measurement
maximizing Eve's information is also optimal from the point of
view of the intrinsic information.
This sounds very plausible. We explore this possibility in
subsection 3.3: for the $(d+1)$-bases and 2-bases protocol with
$d=3$, we construct explicitly the channel $E\rightarrow \bar{E}$
that Eve must apply to her data in order to obtain
$I(A:B|\bar{E})=0$. For the 2-bases protocol and $d=2$, the
channel was given in Ref. \cite{crypto2000}.

\subsection{Step 1: Entanglement distillation}
\label{subed} \noindent We want to study the entanglement
distillation properties of $\rho_{AB}$ for both 2-bases and
$(d+1)$-bases protocols. In order to do that, we first calculate
its partial transposition. It is well known that a state with
positive partial transpose (PPT) is not distillable \cite{bound}.
This would define a critical $D$, denoted by $D^{ED}$, above which
the state cannot be distilled. Moreover, we will see that below
this value the fidelity of $\rho_{AB}$ with a two-qudit maximally
entangled state satisfies \ba \sandwich{\Phi}{\rho_{AB}(F)}{\Phi}&
>&\frac{1}{d}\,.\label{conddist}\ea This condition is sufficient
for distillability \cite{redcrit}. Therefore, $\rho_{AB}$ is
distillable if and only if $D<D^{ED}$, i.e. the non-positivity of
the partial transposition is a necessary and sufficient condition
for the distillability of states (\ref{rhoab}).

\subsubsection{$(d+1)$-bases protocols}
\noindent Inserting (\ref{asd1}) into (\ref{rhoab}), we find that
for the $(d+1)$-bases protocols the state of Alice and Bob after
Eve's attack is simply \ba \rho_{AB}(F)&=&\lambda\,
\ket{\Phi}\bra{\Phi}+(1-\lambda)\,\frac{\one}{d^2}
\label{rhod1}\ea with $\lambda = v^2-z^2 = \frac{dF-1}{d-1}$. The
smallest eigenvalue of the partial transpose $\rho_{AB}^{T_A}$ is
simply
$\lambda_{min}=\lambda(-\frac{1}{d})+(1-\lambda)\frac{1}{d^2}=
\frac{1-(d+1)\lambda}{d^2}$, where $-\frac{1}{d}$ is the minimal
eigenvalue of $(\ket{\Phi}\bra{\Phi})^{T_A}$. The partial
transpose $\rho_{AB}^{T_A}$ is non-negative if $\lambda_{min}\geq
0$, that is if $\lambda\geq\frac{1}{d+1}$ or equivalently
$F\geq\frac{2}{d+1}$. This is precisely the range of value of $F$
for which (\ref{conddist}) does not hold. We have thus proven
that: \ba \mbox{$(d+1)$-bases:  }\; D_{d+1}^{ED}&=&
\frac{d-1}{d+1}\,. \label{dedd1}\ea Moreover, a state of the form
(\ref{rhod1}) cannot be bound-entangled, i.e. the positivity of
its partial transposition is equivalent to separability
\cite{redcrit}.

\subsubsection{2-bases protocols}
\label{subsubed}  \noindent Inserting (\ref{as2}) into
(\ref{rhoab}), and noticing that $x^2=Fy$, we find that for the
2-bases protocols the state of Alice and Bob after Eve's attack is
\ba \rho_{AB}(F)&=&(F^2-y^2)\, \ket{\Phi}\bra{\Phi}
\,+\,y^2\,\one\,+\nonumber\\&& +\, (F-y)y\,\Big(\sum_{m\neq
0}P_{m,0}+ \sum_{n\neq 0}P_{0,n}\Big) \label{rho2}\ea where
$P_{m,n}= \ket{B_{m,n}}\bra{B_{m,n}}$, and recall that
$y=\frac{1-F}{d-1}$. In the computational product basis we have:
\ba
\begin{array}{lcl}
d\,\sandwich{k\,k}{\rho_{AB}(F)}{k\,k}&=&F\\
d\,\sandwich{k\,k'}{\rho_{AB}(F)}{k\,k'}&=&y\\
d\,\sandwich{k\,k}{\rho_{AB}(F)}{k'\,k'}&=&F(F-y)\\
d\,\sandwich{k\,k'}{\rho_{AB}(F)}{j\,j'}&=&y(F-y)\,\delta_{(k-k'),(j-j')}
\end{array} \label{matrixel}\ea where $k,k',j,j'\in\{0,1,..., d-1\}$, $k'\neq k$ and $j\neq
k$. Note that for $F=\bar{F}$ it holds $y=F(F-y)$, that is
$\sandwich{k\,k'}{\rho_{AB}(\bar{F})}{k\,k'}=
\sandwich{k\,k}{\rho_{AB}(\bar{F})}{k'\,k'}$

Condition (\ref{conddist}) is fulfilled for
$F>\bar{F}=\frac{1}{\sqrt{d}}$, so certainly $D^{ED}\geq 1-
\frac{1}{\sqrt{d}}$. Now we should prove that strict equality
holds, by proving that $\rho_{AB}(\bar{F})$ is PPT. For $d=2$,
that is for the entanglement version of the BB84 protocol, the
calculation is particularly simple and it has been proven in
\cite{wolf}. Note that because for two qubits the negativity of
the partial transpose is {\em necessary} and sufficient condition
for entanglement, $\rho_{AB}(\bar{F})$ is also separable. For
$d>3$ we have demonstrated numerically (see Appendix A) that
$\rho_{AB}(\bar{F})$ is indeed PPT. So we can conclude \ba
\mbox{2-bases: }\; D_2^{ED}&=&1- \frac{1}{\sqrt{d}}\,.
\label{ded2}\ea For $d=3,\ldots,15$, we can numerically prove (see
Appendix B) that $\rho_{AB}(\bar F)$ is separable too. Indeed, it
seems very plausible that PPT is a necessary and sufficient for
separability when the states are diagonal in a basis of maximally
entangled states, as it happens for $\rho_{AB}$ (\ref{rhoab}).

\subsection{Step 2: Advantage distillation protocol}
\label{subad} \noindent We turn now to prove that advantage
distillation works for all $D<D^{ED}$. This can be done by
generalizing the advantage distillation protocol described in Ref.
\cite{wolf} for qubits. It works as follows: Alice wants to
establish the secret dit $X$ with Bob. She considers $N$ items of
her list, $\{a_{i_1},...,a_{i_N}\}$, and sends to Bob on the
public channel the list $\{i_1,...,i_N\}$ and the numbers
$\{\tilde{a}_{i_k}\}$ such that $a_{i_k}+\tilde{a}_{i_k}=X$. Bob
takes the corresponding symbols of his list,
$\{b_{i_1},...,b_{i_N}\}$ and calculates
$b_{i_k}+\tilde{a}_{i_k}$. If he finds the same result $Y$ for all
$k$, he notifies to Alice that the dit is accepted; otherwise,
both discard the $N$ symbols. This protocol shows the features
that we discussed for advantage distillation protocols: it
requires two-way communication (Alice must announce and Bob must
confirm), and its yield is very low with increasing $N$. As far as
Eve is concerned, she can only listen to the communication and
compute from her list $\tilde{e}_{i_k}=e_{i_k}+\tilde{a}_{i_k}$.
If Bob accepts, she cannot do better than a majority guess.

Now, recall the purpose we want to achieve: we start in a
situation in which $I(A:E)=I(B:E)$ is larger than $I(A:B)$, and we
want to reverse this situation in order to enter the region in
which the much more efficient one-way protocols can be used. Thus,
we want to show that, after running the above protocol with $N$
sufficiently large, the much shorter lists of dits are such that
Bob's error $\beta_N$ in guessing Alice's dit has become {\em
smaller} than Eve's error $\epsilon_N$ (noted $\gamma_N$ in
\cite{wolf}). So now we must estimate $\beta_N$ and $\epsilon_N$.

Bob accepts a dit when either all his symbols are identical to
those of Alice, which happens with probability $F^N$, or all his
symbols are different from Alice's by the same amount, which
happens with probability $D_{N}=
(d-1)\left(\frac{D}{d-1}\right)^N$. Thus, the probability of Bob
accepting a wrong dit, conditioned to the acceptance, is \ba
\beta_N&=& \frac{D_{N}}{F^N +
D_{N}}\,\leq\,(d-1)\left(\frac{D}{(d-1) \,F}\right)^N\,.
\label{betan}\ea Note that in the limit of large $N$ the previous
expression becomes an equality.

It is more tricky to obtain an estimate for $\epsilon_N$. When Bob
accepts a symbol, Eve makes a majority guess. Of course, there are
enormously many possibilities for Eve to guess wrongly, and it
would be very cumbersome to sum up all of them. The idea is rather
to find those errors that are the most frequent ones. We shall
obtain a bound $\epsilon_N$ which is smaller than the true one,
but very close to it for large $N$ (equal when
$N\rightarrow\infty$). The estimate is based on the following
idea: before the advantage distillation protocol, Eve is strongly
correlated with Alice and Bob. On the one hand, this implies that
when one symbol is more frequent than all the others in Eve's
processed $\tilde{E}$ list, it will almost always be the correct
one. On the other hand, it is very improbable that three or more
symbols appear with the same frequency in the $\tilde{E}$ list.
All in all, the dominating term for Eve's errors should be
associated to the case where {\em two symbols} appear in
$\tilde{E}$ with the same frequency, in which case Eve guesses
wrongly half of the times.

Suppose then that two symbols $x$ and $x'$ appear $M$ times in
$\tilde{E}$, and all the other $d-2$ symbols appear
$M'=\frac{N-2M}{d-2}$. Suppose now that one of the two symbols is
the good one: this is highly probable when $M>M'$, and a situation
in which $M'>M$ is very unlikely to happen. Moreover, we suppose
that $a_{i_k}=b_{i_k}=x$ (the other situation,
$a_{i_k}=b_{i_k}+c=x$, adds only corrections of order $\beta_N$).
The probability that $\tilde{E}$ contains $M$ times $x$ and $x'$
and $M'$ times all the other values is
$\delta^M\left(\frac{1-\delta}{d-1}\right)^{N-M}$ where $\delta$
is the probability that Eve guesses correctly Bob's dit,
conditioned to the fact that Alice's and Bob's dits are equal. As
we said, half of the times Eve will guess $x$ correctly, and half
of the times she will guess $x'$ wrongly. Adding the combinatorial
factor that counts all the possible ways of distributing $x$ and
$x'$ among the $d$ symbols we obtain the estimate \ba \epsilon_N
&\geq & \demi\,{\sum_{M=0}^{N/2}}\frac{N!}{(M!)^2\,\left[
\big(\frac{N-2M}{d-2}\big)! \right]^{d-2}}\, \delta^M\,
\left(\frac{1-\delta}{d-1}\right)^{N-M} \ea and applying
Stirling's approximation $(x!)^m\simeq \frac{(mx)!}{m^{mx}}$ we
find the asymptotic behavior \ba \epsilon_N &\geq & k\,
\left(2\sqrt{\delta\frac{1-\delta}{d-1}}+
(d-2)\frac{1-\delta}{d-1}\right)^{N} \label{epsn} \ea with $k$
some positive constant. Comparing this expression with
(\ref{betan}), we see that $\beta_N$ decreases exponentially
faster than $\epsilon_N$ whenever \ba \frac{D}{(d-1) \,F}\;&< &
\;2\sqrt{\delta\frac{1-\delta}{d-1}}+ (d-2)\frac{1-\delta}{d-1}
\,. \label{ineqfin}\ea The value of $\delta$ is found reading
through Ref. \cite{cerf}. For the 2-bases protocol, the
probability that Eve guesses correctly is independent of the
correlation Alice-Bob, so $\delta_2=F_E$ given by \ba
F_E=\frac{F}{d}+\frac{(d-1)(1-F)}{d}+\frac{2}{d}\sqrt{F(1-F)(d-1)}
\,.\ea For the $(d+1)$-bases protocols, $\delta_{d+1}=
(F+F_E-1)/F$, where $F_E=1-\frac{d-1}{d}(v-z)^2$. Inserting these
values into (\ref{ineqfin}), we find after some algebra that the
condition is satisfied precisely for $D<D^{ED}$ given by
(\ref{ded2}), resp. (\ref{dedd1}). Thus, our advantage
distillation protocol works at least up to $D^{ED}$.

\subsection{Intrinsic information at $D=D^{ED}$ for $d=3$}
\label{subii} \noindent In this subsection, we want to prove that
the intrinsic information (\ref{intrinf}) of $P(A,B,E)$ goes to
zero at $D=D^{ED}$, when Eve applies the measurements of Ref.
\cite{cerf}. As said above, this quantity provides an upper bound
for the amount of secret bits the honest parties can extract from
a probability distribution. Since $\rho_{AB}$ at $D=D^{ED}$ is
separable, we already know that there exists a measurement for Eve
such that $I(A:B\downarrow E)=0$ for all Alice's and Bob's
measurements \cite{crypto2000}. Thus, the state is completely
useless for establishing a key. Here, we study whether the
measurements maximizing Eve's mutual information are also optimal
from the point of view of the intrinsic information, when
$D=D^{ED}$. We shall give the complete proof only for $d=3$, but
we start with general considerations.

After basis reconciliation, Alice, Bob and Eve share the
probability distribution $P(a,b,e_a,\chi)$, that can be found
reading through Ref. \cite{cerf} --- recall that $\chi=b-a$
deterministically. For the 2-bases protocol, we have: \ba
\begin{array}{lcl}
P(a,b=a,e_a=a,0)&=&F\,F_E/d\\
P(a,b=a,e_a\neq a,0)&=&F\,D_E/d\\
P(a,b\neq a,e_a=a,b-a)&=&D\,F_E/d\\
P(a,b\neq a,e_a\neq a, b-a)&=&D\,D_E/d\,.
\end{array}
\ea For $(d+1)$-bases protocols, writing $\lambda=(F+F_E-1)/F$, we
have: \ba
\begin{array}{lcl}
P(a,b=a,e_a=a,0)&=&F\,\lambda/d\\
P(a,b=a,e_a\neq a,0)&=&F\,(\frac{1-\lambda}{d-1})/d\\
P(a,b\neq a,e_a=a,b-a)&=&D/d\\
P(a,b\neq a,e_a\neq a, b-a)&=&0\,.
\end{array}
\ea For both these distributions, the conditional mutual
information is $I(A:B|E)\neq 0$. We are looking for a classical
channel $\cal{C}$ that Eve could apply to her information \ba
{\cal{C}}\,:\,E=\{(e_a,\chi)\}&\rightarrow & \bar{E}= \{\bar{u}\}
\ea in such a way that $I(A:B|\bar{E})= 0$ \cite{note9}. The
channel is defined by the probabilities $C(\bar{u}|e_a,\chi)$ that
the symbol $(e_a,\chi)$ of $E$ is sent onto the symbol $\bar{u}$
of $\bar{E}$. Of course, these probabilities fulfill the condition
$\sum_{\bar{u}}C(\bar{u}|e_a,\chi)=1$. The new probability
distribution for Alice, Bob and Eve is given by \ba
P(a,b,\bar{u})&=&\sum_{e_a,chi}C(\bar{u}|e_a,\chi)
P(a,b,e_a,\chi)\,,\ea whence conditional probabilities
$P(a,b|\bar{u})$ are obtained in the usual way.

At this stage, we know of no systematic way of finding the channel
that minimizes $I(A:B|\bar{E})$, so we shall try to describe our
intuition. Basically, one must keep in mind that
$I(A:B|\bar{E})=0$ if and only if $P(a,b|\bar{u})$ is in fact the
product probability $P(a|\bar{u})P(b|\bar{u})$. In particular,
identities like \ba P(a,b|\bar{u})P(a',b'|\bar{u})&=&
P(a,b'|\bar{u})P(a',b|\bar{u}) \ea should hold for all values of
the symbols.

For $d=3$, we tried the ``simplest" form of the channel and
verified that it gives indeed $I(A:B|\bar{E})=0$ for $D=D^{ED}$.
It is defined as follows:

\begin{itemize}
\item The symbol $\bar{E}$ is a trit:
\ba \bar{E}&=&\{u_0,u_1,u_2\}\,. \ea
\item When Eve has introduced no error ($\chi=0$), Eve's guess is
sent deterministically on the corresponding value of the trit: \ba
C(u_k|e_a,\chi=0)&=&\delta_{k,e_a}\,. \ea
\item When Eve has introduced some errors, Eve's guesses are mixed
according to the following rule: \ba C(u_k|e_a,\chi\neq 0)&=&
\begin{array}{lcl}
c&,&k\neq e_a-\chi\\ 1-2c&,&k= e_a-\chi
\end{array}\,. \ea
\end{itemize}
The value of the parameter $c$ was found on the computer. For the
2-bases protocol, we found $c\approx 0.4715$; for the 4-bases
protocol, $c\approx 0.4444$.

\section{The CK bound and the violation of Bell's inequalities}
\noindent As we said, although
strictly speaking a secret key can be extracted for $D<D^{AD}$, in
practice the extraction can be made {\em efficiently} only for
$D<D^{CK}$, and this criterion is the most studied in the
literature. The value of $D^{CK}$ for the protocols we are
considering is given in Ref. \cite{cerf}. For 2-bases protocols,
$D_2^{CK}\,=\,\demi\,\big(1- \frac{1}{\sqrt{d}}\big)\,=\,\demi
D_2^{AD}$. For the $(d+1)$-bases protocols, it is cumbersome to
give a closed formula for $D_{d+1}^{CK}$, but it is slightly
larger than $D_2^{CK}$: in other words, $(d+1)$-bases protocols
are more robust than 2-bases protocols also if one considers the
CK bound.

We saw in the previous Section that $D^{AD}=D^{ED}$: advantage
distillation is tightly linked to entanglement distillation.
According to this intuition, one expects $D^{CK}$ to be linked to
entanglement distillation using one-way communication \cite{BDSW}.
As far as we know, there are few results in this direction.
Remarkably, the bound $D^{CK}$ also seems to be linked with the
{\em violation of a Bell's inequality}, but it is unclear whether
this link is as tight as (\ref{thm1}), because it is a hard
problem to characterize all the Bell's inequalities. More
precisely, the state-of-the-question is described by the following

Statement: {\em Define the two bounds: (i)
$I(A:B)>\min\big[I(A:E),I(B:E)\big]$ for $D<D^{CK}$, and (ii)
$\rho_{AB}(F)$ violates a Bell's inequality for $D=1-F<D^{Bell}$.
Then, for any $d$, for both the 2-bases and the the $(d+1)$-bases
protocols, and for all known Bell inequalities, it holds} \ba
D^{Bell}&\leq&D^{CK}\,.\label{thm2}\ea In words: if the state
$\rho_{AB}$ violates a Bell's inequality, then certainly the
correlations can be used to extract a secret key in an efficient
way. This is one of the situations in which Bell's inequalities
show themselves as {\em witnesses of useful entanglement}.

We start with a review of the $d=2$ case. 
%
Consider first the 2-bases protocol. Writing as usual
$\ket{\Phi^{\pm}}= \frac{1}{\sqrt{2}}(\ket{00}\pm\ket{11})$ and
$\ket{\Psi^{\pm}}= \frac{1}{\sqrt{2}}(\ket{01}\pm\ket{10})$, the
state (\ref{rho2}) becomes $F^2
P_{\Phi^+}+F(1-F)\big[P_{\Phi^-}+P_{\Psi^+}\big]+(1-F)^2
P_{\Psi^-}$, that is \ba
\rho_{AB}(F)&=&\frac{1}{4}\,\Big(\one+\sum_{k=x,y,z}t_k(F)\,\si_k\otimes
\si_k \Big) \label{rho2qubits}\ea with $t_x=t_z=2F-1$ and
$t_y=-(2F-1)^2$. Applying the Horodeckis' result \cite{horo}, the
expectation value for the CHSH-Bell operator \cite{chsh} with the
optimal settings is given by
$S=\sqrt{t_x^2+t_z^2}=(2F-1)\sqrt{2}$. The Bell inequality is
violated for $S>1$, that is for $F>\demi(1+\frac{1}{\sqrt{2}})$,
that is again for $D<D^{Bell}=\demi(1-\frac{1}{\sqrt{2}})=D^{CK}$.
So for the qubit protocol the equality holds in (\ref{thm2}).

This seems to be no longer true when we move to the 3-bases
protocol (six-states protocol). The state (\ref{rhod1}) has the
same form as (\ref{rho2qubits}), with $t_x=t_z=-t_y=2F-1$. The
condition for the violation of the CHSH-Bell inequality is then
exactly the same as before, so we find again
$D^{Bell}=\demi(1-\frac{1}{\sqrt{2}})$. But for the six-states
protocol, the bound $D^{CK}$ is slightly larger than this value.

One might start questioning the choice of the inequality. In the
CHSH inequality \cite{chsh}, Alice and Bob choose each among {\em
two} possible settings. For this reason, the inequality seems
suited for the 2-bases protocol (although the settings are not the
same ones), while for the 3-bases protocol one should find an
inequality with three settings per qubit. Recently, the complete
characterization of all the inequalities with three settings of
two outcomes per side has been achieved \cite{Sliwa,CG2}. None of
these inequalities fills the gap between $D^{CK}$ and $D^{Bell}$.

Moving now to the $d>2$ case, the knowledge is even more vague.
Good Bell's inequalities for two entangled qudits for $d>2$ have
been found only recently \cite{collins,helle}. When applied to our
problem, all these inequalities give $D^{Bell}<D^{CK}$ both for
the 2-bases and the $(d+1)$-bases protocols. Note that the
inequality with two settings per qudit of Collins et al.
\cite{collins} is in some sense optimal \cite{CG2,lluis}.

\section{Concluding remarks}
\noindent In this article we have studied the relation between
quantum and classical distillation protocols for quantum
cryptography. We have shown that classical and quantum key
distillation protocols work up to the same point or disturbance
for the schemes using two and $d+1$ bases, when individual attacks
based on cloning machines are considered. Indeed, this equivalence
has been recently extended in Ref. \cite{AMG} to all two-qubit
entangled states, and therefore to all the so-called one-copy
distillable states (which include the states studied in this
article), and to all individual attacks. We would like to conclude
the present work with a list of several open questions connected
to many of the points raised here. The solution of any of them
will provide more insight into the relation between classical and
quantum distillation protocols for quantum key distribution.

\begin{itemize}
\item The first open question concerns of course the validity of
our results when some of the assumptions made for Eve are relaxed.
Although these assumptions seem very reasonable taking into
account present-day technological limitations, they are quite
strong from a theoretical point of view. First, one may wonder
what happens if Eve changes her attack, still individual, from
symbol to symbol. In this more general scenario, the so-called
collision probability provides the honest parties with a bound on
the amount of privacy amplification needed for distilling a secure
key \cite{BBCM,Lutkenhaus}. One can also consider collective
attacks where Eve interacts with more than one qudit \cite{Mor}.
Or even if the interaction is done symbol by symbol, she may delay
her final measurement until the end of the classical communication
between the honest parties \cite{notesc}. In all these situations,
the eavesdropper is more powerful than in this work, so they
clearly deserve further investigation. \item Another open question
is the validity of the conjecture that the cloning machines
defined above provide really the optimal individual eavesdropping,
also for $d>3$. While this seems very plausible for the
$(d+1)$-bases protocols, also when the Theorem (\ref{thm1}) of
this paper is taken into account, some doubts can be raised for
the 2-bases protocols. In these protocols, the second basis has
always been defined as the Fourier-dual basis of the computational
basis. For $d=2$ and $d=3$ this is not a restriction, since the
following holds: for any $B_1$, $B_2$ and $B_3$ mutually maximally
conjugated bases, there exist a unitary operation that sends the
pair $(B_1,B_2)$ onto the pair $(B_1,B_3)$. For eavesdropping on
QC, this means that the cloning machines $C_{12}$ and $C_{13}$
that are optimized for, respectively, $(B_1,B_2)$ and $(B_1,B_3)$,
are equivalent under a unitary operation, so in particular have
the same fidelity and define the same bounds. For $d>3$ however,
it is in general impossible to link $(B_1,B_2)$ to $(B_1,B_3)$
with a unitary operation \cite{related}. This opens some
intriguing possibilities: for instance, it might turn out that
some pairs of mutually conjugated bases are more difficult to
clone than others, and are therefore more suitable for
cryptography. Recent results \cite{related2} suggest that this may
not be the case and that all pairs of mutually conjugated bases
may be equivalent for quantum cryptography, although this is still
an open question. \item A related open question concerns the
choice of Eve's strategy. As mentioned explicitly, we have always
supposed in this paper --- as is done, to our knowledge, in most
of the papers on QC --- that Eve's best individual attack is the
one that maximizes Eve's information at any given error rate
induced on the correlations Alice-Bob. But Eve might have a
different purpose; for instance, since after all the security of
QC cannot be beaten, she might be willing to {\em decrease the
robustness}. Thus, she may decide to apply the attack that
introduces the minimal disturbance and lowers the intrinsic
information of the resulting probability distribution. This is
also connected to the security of quantum channels. Indeed, from
the cryptography point of view, Eve's attack completely defines a
channel. Therefore, when does a given channel allow for a secure
key distribution, assuming that all the errors are due to the
presence of an eavesdropper? Recent results in \cite{AMG} suggest
that only those channels that allow to distribute distillable
entanglement are secure. \item The last question deals with more
quantitative aspects. In Section 3, we have shown that two
protocols for extracting a secret key, namely ``measurement
followed by advantage distillation" and ``entanglement
distillation followed by measurement", work up to the same error
rate. However, one of these two strategies might turn out to have
a better {\em yield} than the other one. This is a complicated
problem since, for both advantage distillation and entanglement
distillation, the optimal protocols are not known.
\end{itemize}

\bigskip

{\em Note added in proof:} The same results as in section 3 have
been simultaneously and independently found in Ref. \cite{BCEEKM}.
There, the analysis is restricted to $d+1$-bases protocols.

\section{Acknowledgements}
\noindent We thank Renato Renner and Stefan Wolf for discussion,
and Serge Massar, Jonathan Barrett and an anonymous referee for
insightful comments on the first version. We acknowledge financial
supports by the Swiss OFES and NSF within the European project
RESQ (IST-2001-37559) and the Swiss national center ``Quantum
Photonics".

\section{Appendix A}
\noindent In this Appendix, we describe the
efficient numerical calculation used to demonstrate that
$\rho_{AB}(\bar{F})$ for the 2-bases protocol is PPT (see
paragraph \ref{subsubed}).

When one resorts to numerical methods, the first idea would be to
use the brute force of the computer: write a program that takes
$\rho_{AB}(\bar{F})\equiv\rho$, computes $\rho^{T_A}\equiv M$ and
finds its minimal eigenvalue. But $M$ is a $d^2\times d^2$ matrix,
and since it has a nice structure one can do much better.
Actually, we show below that $M$ is actually block-diagonal, with
$d$ blocks of dimension $d\times d$. For odd $d$, all the blocks
are identical; for even $d$, two different blocks appear, each in
$\frac{d}{2}$ copies. Having noticed that, one has to find
numerically the minimal eigenvalue of one or two $d\times d$ real
matrices, and this scales much better than the brute force method.
Based on this result, we could very easily check that
$\rho_{AB}(\bar{F})$ is PPT up to $d=200$, this number having no
other meaning than the fact that one must stop the computation
somewhere --- anyway, it is unlikely that a QC protocol using
entangled states of two 200-levels systems will ever be of any
practical interest.

To study the structure of $M=\rho_{AB}^{T_A}$, we take the partial
transpose of (\ref{matrixel}): \ba
\begin{array}{lcl}
\sandwich{k\,k}{M}{k\,k}&=&A\\
\sandwich{k\,k'}{M}{k\,k'}&=&B\\
\sandwich{k\,k'}{M}{k'\,k}&=&B'\\
\sandwich{k\,k'}{M}{j\,j'}&=&C\,\delta_{(k+k'),(j+j')}
\end{array} \label{matrel2}\ea
with $A=\frac{F}{d}$, $B=\frac{y}{d}$, $B'=\frac{F(F-y)}{d}$ and
$C=\frac{y(F-y)}{d}$. Recall that $B=B'$ for $F=\bar{F}$; we must
prove that the minimal eigenvalue of $M$ is negative if and only
if $B<B'$. From (\ref{matrel2}) it is then clear that $M$ is
composed of $d$ blocks $d\times d$, because these four relations
show that only the $\sandwich{k\,k'}{M}{j\,j'}$ with $k+k'=j+j'$
are non-zero. Explicitly, defining the vector ${\bf{c}}=\big(C\;C
\big)$ and the $2\times 2 $ blocks \ban \bf{A}=\left(\begin{array}{cc}A & C\\
C&A
\end{array}\right)\,,&\, \bf{B} = \left(\begin{array}{cc}B & B'\\ B'& B \end{array}\right)
\,,&\, \bf{C}= \left(\begin{array}{cc}C & C\\ C& C
\end{array}\right) \ean
one finds the following structure for $M$:\\ odd $d$: all blocks
are identical to \ba
\left(\begin{array}{ccccc} A & \bf{c} & \bf{c} &  ... & \bf{c} \\
\bf{c}^T & \bf{B} & \bf{C}  & ... & \bf{C} \\
\bf{c}^T & \bf{C} & \bf{B} & ... & \bf{C}\\
\vdots & & & \ddots  & \vdots\\
\bf{c}^T & \bf{C} & \bf{C} & ... & \bf{B}
\end{array}\right)\,; \ea
even $d$: the $\frac{d}{2}$ blocks characterized by $k+k'$ even
are equal to \ba
\left(\begin{array}{ccccc} \bf{A} & \bf{C} & \bf{C} & ... &  \bf{C} \\
\bf{C} & \bf{B} & \bf{C}  & ... & \bf{C} \\
\bf{C} & \bf{C} & \bf{B} & ... & \bf{C}\\
\vdots & & & \ddots  & \vdots\\
\bf{C} & \bf{C} & \bf{C} & ... & \bf{B}
\end{array}\right)\,; \ea
the $\frac{d}{2}$ blocks characterized by $k+k'$ odd are equal to
\ba
\left(\begin{array}{ccccc} \bf{B} & \bf{C} & \bf{C} & ... &  \bf{C} \\
\bf{C} & \bf{B} & \bf{C}  & ... & \bf{C} \\
\bf{C} & \bf{C} & \bf{B} & ... & \bf{C}\\
\vdots & & & \ddots  & \vdots\\
\bf{C} & \bf{C} & \bf{C} & ... & \bf{B}
\end{array}\right)\,. \ea
So these are the $d\times d$ matrices whose minimal eigenvalue is
to be found.

\section{Appendix B}
\noindent In this appendix we show how to
numerically prove the separability of the states $\rho_{AB}(\bar
F)$ for the 2-bases protocol. Note that all the states
$\rho_{AB}(\bar F)$ are diagonal in the Bell basis
$\{\ket{B_{m,n}}\}$ (\ref{rhoab}). This turns out to be the
crucial point in our demonstration. Indeed, it is very plausible
that PPT is a necessary and also sufficient condition for the
separability of Bell diagonal states, but we are not aware of any
proof of that.

Any density matrix, $\rho$, can be brought into a Bell diagonal
form by a sequence of local operations assisted with classical
communication (LOCC). This is done by the following depolarization
protocol \ba {\mathcal D}(\rho)&=&\sum_{m,n}
\frac{1}{m\,n}(U_{m,n}\otimes U_{m,n}^*)\rho(U_{m,n}\otimes
U_{m,n}^*)^\dagger\,,\ea that makes the transformation \ba
{\mathcal D}(\rho)&\longrightarrow&
\sum_{m,n}\lambda_{m,n}\ket{B_{m,n}}\bra{B_{m,n}}\,,\ea where
$\lambda_{m,n}=\bra{B_{m,n}}\rho\ket{B_{m,n}}$. Thus, the overlaps
with the Bell states for the initial and the depolarized state are
the same, they are not changed by ${\mathcal D}$.

We consider a subset of the set of separable pure states in
$\compl^d\otimes\compl^d$ parameterized as \ba
\ket{\psi_s}=\ket{\psi}\otimes\ket{\psi^*}\,.\ea Note that these
states depend on $2d-2$ parameters, instead of the $2(d-2)$ needed
for a generic separable pure state. We look for those
$\ket{\psi_s}$ minimizing the function \ba
f(\psi_s)=\sum_{m,n}(|a_{m,n}(\bar
F)|^2-|\bra{B_{m,n}}\psi_s\rangle|^2)^2\,. \ea After some computer
runs, we always find (up to $d=15$) a state $\ket{\bar\psi_s}$
such that $f(\bar\psi_s)\simeq 0$, which means that
$|\bra{B_{m,n}}\bar\psi_s\rangle|\simeq |a_{m,n}(\bar F)|$.
Therefore, after applying the depolarization protocol to this
state, one obtains \ba \rho_{AB}(\bar F)\simeq {\mathcal
D}(\ket{\bar\psi_s}\bra{\bar\psi_s})\,,\ea which means that
$\rho_{AB}(\bar F)$ is separable.


\begin{thebibliography}{000}

\bibitem{review}
N. Gisin, G. Ribordy, W. Tittel, H. Zbinden, Rev. Mod. Phys {\bf
74}, 145 (2002); {\em The Physics of Quantum Information}, ed. by
D. Bouwmeester, A. Ekert and A. Zeilinger, Springer-Verlag (2000).
\bibitem{uncond}
This argument has to be understood in a rather qualitative way.
Proofs of unconditional security are much more involved, see for
instance D. Mayers, quant-ph/9802025; H.-K. Lo and H. F. Chau,
Science {\bf 283}, 2050 (1999); P. W. Shor and J. Preskill, Phys.
Rev. Lett. {\bf 85}, 441 (2000).
\bibitem{csi} I. Csisz\'ar and J. K\"{o}rner, IEEE Trans. Inf. Theory
{\bf IT-24}, 339 (1978).
\bibitem{BBCM}
C. H. Bennett, G. Brassard, C. Cr\'epeau and U. Maurer, IEEE
Trans. Inf. Theory {\bf 41}, 1915 (1995).
\bibitem{maurer} U.M. Maurer, IEEE Trans. Inf. Theory
{\bf 39}, 733 (1993).
\bibitem{wolf} N. Gisin and S. Wolf, Phys. Rev. Lett. {\bf 83}, 4200
(1999).
\bibitem{cerf} N.J. Cerf, M. Bourennane, A. Karlsson and N. Gisin,
Phys. Rev. Lett. 88, 127902 (2002).
\bibitem{bell1} B. Huttner and N. Gisin, Phys. Lett. A {\bf 228}, 13
(1997).
\bibitem{fuchs} C. Fuchs, N. Gisin, R.B. Griffiths, C.-S.
Niu and A. Peres, Phys. Rev. A {\bf 56}, 1163 (1997).
\bibitem{BB84}
C. H. Bennett and G. Brassard, in {\em Proceedings IEEE Int. Conf.
on Computers, Systems and Signal Processing, Bangalore, India}
(IEEE, New York, 1984), pp. 175-179.
\bibitem{wootters} The existence of $d+1$ bases has been
(constructively) demonstrated only when $d$ is a power of a prime
number: W.K. Wootters and B.D. Fields, Ann. Phys. (N.Y.) {\bf
191}, 363 (1989).
\bibitem{BBM}
C. H. Bennett, G. Brassard and N. D. Mermin, Phys. Rev. Lett. {\bf
86}, 557 (1992).
\bibitem{noteekert}
The use of two-qubit entangled states for a secure key
distribution was originally proposed in A. Ekert, Phys. Rev. Lett.
{\bf 67}, 661 (1991). There, Eve's intervencion is detected by
checking the violation of a Bell inequality.
\bibitem{CG}
J. I. Cirac and N. Gisin, Phys. Lett. A {\bf 229}, 1 (1997).
\bibitem{exist}
Note that it is assumed the (unproven) existence of the $d+1$
maximally conjugated bases for all dimension.
\bibitem{sixstate} D. Bruss, Phys. Rev. Lett. {\bf 81}, 3018 (1998);
H. Bechmann-Pasquinucci and N. Gisin, Phys. Rev. A {\bf 59}, 4238
(1999).
\bibitem{bruss} D. Bruss and C. Macchiavello, Phys. Rev. Lett. 88, 127901
(2002).
\bibitem{clon}
N. J. Cerf, Phys. Rev. Lett. {\bf 84}, 4497 (2000); J. Mod. Opt.
{\bf 47}, 187 (2000); Acta Phys. Slov. {\bf 48}, 115 (1998).
\bibitem{chefles}
A. Chefles, Contemp. Phys. {\bf 41}, 401 (2000) and references
therein.
\bibitem{notepr}
In order to transform the quantum state $\ket{\Psi}_{ABE}$ into
the probability distribution $P(A,B,E)$, Eve's measurements should
be specified. As said in section 2, we consider the measurements
of Ref. \cite{cerf}, designed for maximizing Eve's information.
\bibitem{crypto2000} N. Gisin and S. Wolf, {\em Proceedings of CRYPTO
2000}, Lecture Notes in Computer Science {\bf 1880}, 482,
Springer-Verlag, 2000.
\bibitem{bound}
M. Horodecki, P. Horodecki and R. Horodecki, Phys. Rev. Lett. {\bf
80} 5239 (1998).
\bibitem{redcrit}
M. Horodecki and P. Horodecki, Phys. Rev. A {\bf 59}, 4206 (1999).
\bibitem{note9} 
In general, the vanishing of intrinsic information does not imply
the existence of a channel $\cal{C}$: strictly speaking, it would
be enough to have a sequence of channels ${\cal{C}}_n$ such that
the intrinsic information converges to 0 when $n$ goes to
infinity.
\bibitem{BDSW}
C. H. Bennett, D. P. DiVincenzo, J. A. Smolin and W. K. Wootters,
Phys. Rev. A {\bf 54}, 3824 (1996).
\bibitem{horo} M. Horodecki, P. Horodecki and M. Horodecki, Phys. Lett. A {\bf
200}, 340 (1995).
\bibitem{chsh} J. F. Clauser, M. A. Horne, A. Shimony and R. A.
Holt, Phys. Rev. Lett. {\bf 23}, 880 (1969).
\bibitem{Sliwa} C. Sliwa, quant-ph/0305190.
\bibitem{CG2} D. Collins and N. Gisin, quant-ph/0306129.
\bibitem{collins} D. Collins, N. Gisin, N. Linden, S. Massar and S.
Popescu, Phys. Rev. Lett. {\bf 88}, 040404 (2002).
\bibitem{helle} H. Bechmann-Pasquinucci and N. Gisin,
Phys. Rev. A {\bf 67}, 062310 (2003).
\bibitem{lluis} Ll. Masanes, Quant. Inf. Comp. {\bf 3},
No. 4, 345 (2003).
\bibitem{AMG}
A. Ac\'\i n, Ll. Masanes and N. Gisin, Phys. Rev. Lett. 
{\bf 91}, 167901 (2003)
\bibitem{Lutkenhaus}
N. L\"utkenhaus, Phys. Rev. A {\bf 54}, 97 (1996).
\bibitem{Mor}
See, for instance, T. Mor, PhD Thesis, April 97, Technion, Haifa,
Israel, also available as quant-ph/9906073.
\bibitem{notesc}
This type of attacks is not entirely individual, in spite of the
fact that the interaction is done symbol by symbol, because the
final measurement may include a global entangling operation among
all the qudits kept by Eve.
\bibitem{related}
Some results in this direction can be found in A. Wojcik, A.
Grudka and R. W. Chhajlany, quant-ph/0305034
\bibitem{related2}
T. Durt and B. Nagler, Phys. Rev. A {\bf 68}, 042323 (2003) .
\bibitem{BCEEKM} D. Bruss, M. Christandl, A. Ekert, B.-G. Englert,
D. Kaszlikowski and C. Macchiavello, Phys. Rev. Lett. 
{\bf 91}, 097901 (2003).


\end{thebibliography}
\end{document}